\documentclass[11pt, a4paper]{article}

\usepackage[margin=1in]{geometry}
\usepackage{tabularx}
\newcolumntype{Y}{>{\centering\arraybackslash}X}
\usepackage{graphicx}
\usepackage{booktabs}
\usepackage{multirow}
\usepackage[table]{xcolor}
\usepackage{xcolor}
\usepackage{float}
\usepackage{amsmath,amssymb,amsfonts,amsthm}
\usepackage{mathrsfs}
\usepackage{changes}
\usepackage{textcomp}
\usepackage[version=4]{mhchem}
\usepackage[T1]{fontenc}
\usepackage[utf8]{inputenc}
\usepackage{authblk}
\usepackage{fontawesome5}
\usepackage[colorlinks=true,linkcolor=blue,citecolor=blue,urlcolor=blue]{hyperref}

\usepackage{algorithm}
\usepackage{algorithmicx}
\usepackage{algpseudocode}
\usepackage{listings}

\theoremstyle{plain}

\theoremstyle{definition}

\newcommand{\bibcommenthead}{}
\begin{document}

\title{QUASAR: A Universal Autonomous System for Atomistic Simulation and a Benchmark of Its Capabilities}

\author[1]{Fengxu Yang}
\author[1]{Jack D. Evans\thanks{Corresponding author: j.evans@adelaide.edu.au}}

\affil[1]{School of Physics, Chemistry and Earth Sciences, Adelaide University, North Terrace, Adelaide, 5005, South Australia, Australia}

\date{}

\maketitle

\begin{abstract}
The integration of large language models (LLMs) into materials science offers a transformative opportunity to streamline computational workflows, yet current agentic systems remain constrained by rigid, carefully crafted domain-specific tool-calling paradigms and narrowly scoped agents. In this work, we introduce QUASAR, a universal autonomous system for atomistic simulation designed to facilitate production-grade scientific discovery. QUASAR autonomously orchestrates complex multi-scale workflows across diverse methods, including density functional theory, machine learning potentials, molecular dynamics, and Monte Carlo simulations. The system incorporates robust mechanisms for adaptive planning, context-efficient memory management, and hybrid knowledge retrieval to navigate real-world research scenarios without human intervention. We benchmark QUASAR against a series of three-tiered tasks, progressing from routine tasks to frontier research challenges such as photocatalyst screening and novel material assessment. These results suggest that QUASAR can function as a general atomistic reasoning system rather than a task-specific automation framework. They also provide initial evidence supporting the potential deployment of agentic AI as a component of computational chemistry research workflows, while identifying areas requiring further development.
\end{abstract}

\newpage

\section{Introduction}\label{Introduction}
The rapid advancement of agentic capabilities in large language models (LLMs) presents a compelling opportunity to transform computational chemistry for materials science research \cite{wei_ai_2025}. As a discipline fundamentally driven by complex computer-mediated workflows, computational chemistry tasks are naturally suited for integration with LLM-based automation, which can streamline tasks such as simulation setup, data interpretation, and error handling. This approach reduces both cognitive load and the technical burden associated with specialized software stacks and continuous human oversight. More importantly, it can serve as a low-risk testing ground for systems that manage self-driving labs \cite{tom_self-driving_2024, inizan_system_2025}. 

Recent research demonstrates the effective integration of LLMs across multiple stages of computational chemistry pipelines. An early proof of concept, ProtAgents, demonstrated a foundational framework for autonomous simulation platforms. It coordinated multiple AI agents with distinct domain-specific roles, including a planner, a function executing assistant, and a critic. Together, these agents tackled de novo protein design tasks such as structure generation via Chroma, secondary structure analysis, physics based vibrational frequency calculations, and mechanical property prediction using a fine-tuned transformer model.\cite{ghafarollahi_protagents:_2024} Building upon this, Wang et al. developed DREAMS, a hierarchical multi-agent framework for density functional theory (DFT) materials simulations that combines a central LLM planner with domain-specific agents for atomistic structure generation, systematic convergence testing, HPC scheduling, and error handling \cite{wang_dreams_2025}. Similarly, Vriza et al. presented a framework for end-to-end atomistic simulations, using LLMs and specialized agents to automate processes from structure generation via Atomsk and interatomic potential discovery, to MD execution with LAMMPS and analysis using OVITO and Phonopy \cite{vriza_multi-agentic_2026}. Several other frameworks have also explored similar approaches, highlighting the rapidly expanding ecosystem of intelligent, tool-augmented agentic simulation workflows \cite{pham_chemgraph_2025, ghafarollahi_automating_2025, mendible-barreto_dynamate_2025, nduma_crystalyse_2025}.

Most agentic systems for computational chemistry have relied largely on human-crafted scaffolding, including domain-specific tools, fine-grained agent decompositions, and rigid workflows. This was a practical response to the limitations of earlier LLMs, whose capacity for autonomous planning and reasoning was insufficient to be trusted without tight constraints. However, this design philosophy has not kept pace with the rapid progress of the underlying models. Recent advances in agentic applications such as AutoResearch \cite{andrej_karpathy/autoresearch_2026}, show substantial improvements of LLMs in handling complex, open-ended research tasks. Therefore, continuing to impose heavily engineered scaffolding on models risks underutilising their current capabilities.

A major limitation of scaffolding-heavy designs is the implicit assumption that LLMs require human-defined functions to perform domain-specific operations. For instance, dedicated tools for reading or generating crystal structures, or wrappers that invoke simulations with fixed parameter sets. This assumption underestimates the breadth of knowledge already encoded in capable LLMs, which have been exposed to a wide range of computational chemistry techniques, file formats, simulation protocols, and domain conventions during training. Rather than relying on hardcoded interfaces, these models can reason directly about how to construct, interpret, or manipulate domain-relevant representations as the task requires. More importantly, rigid human-defined functions carry inherent scope risks, as they are designed for anticipated use cases and may behave inappropriately when confronted with scenarios beyond their original specification. By contrast, LLMs can natively reason over a problem, adapting their approach to the specifics of each case, handling edge cases, or unconventional workflows without additional engineering effort.

Moreover, introducing a large number of highly specialized agents increases the overall system complexity and can lead to fragile operations, particularly in basic routing, coordination, and control logic, where even minor mismatches in expected behavior can cascade into larger errors. Together, these factors cause substantial engineering burdens, requiring careful manual design, continuous maintenance, and considerable human effort to stabilize and streamline workflows. This complexity ultimately undermines scalability, reliability, and adaptability of agentic systems in real-world computational chemistry settings.

\begin{table}[H]
    \centering
    \label{tab:system_comparison}
    \caption{Comparison of QUASAR with recent agentic computational chemistry systems. SE denotes semi-empirical methods.}

    \vspace{0.3cm}
    \renewcommand{\arraystretch}{1.5}
    
    \newcolumntype{Y}{>{\raggedright\arraybackslash}X}

    \small 
    \setlength{\tabcolsep}{3pt} 

    \begin{tabularx}{\textwidth}{ l >{\columncolor{gray!10}}Y Y Y Y Y }
        \noalign{\hrule height 1.5pt}
        \textbf{Category} &
        \textbf{QUASAR} &
        \textbf{El Agente Quantur} \cite{perez-sanchez_agente_2026} &
        \textbf{DREAMS} \cite{wang_dreams_2025} &
        \textbf{LAMMPS-Agents} \cite{vriza_multi-agentic_2026} \\
        \noalign{\hrule height 1.5pt}

        Domain &
        DFT, MD, MC, MLP &
        DFT, SE &
        DFT &
        MD, MLP \\
        \hline
        
        Agent Count &
        3 &
        58 &
        3 &
        11 \\
        \hline

        Domain-Specific Tools &
        No &
        Yes &
        Yes &
        Yes \\
        \hline
        
        Extensibility &
        On-the-fly tool integration via natural language &
        Modular addition of domain-specific agents and tools &
        Modular addition of domain-specific agents and tools &
        Modular addition of domain-specific agents and tools \\
        \hline

        Benchmark Depth &
        Novel research questions &
        Expert-level challenges  &
        Expert-level challenges  &
        Expert-level challenges  \\
        \hline
        
        Deployment &
        Native Docker; Platform-agnostic  &
        Closed Source; Commercial Cloud &
        Manual setup; Specific HPC only &
        Native Docker; Specific HPC only \\
        \hline

        \noalign{\hrule height 1pt}
    \end{tabularx}
\end{table}

In this work, we introduce QUASAR, a production-level universal atomistic computation system designed to overcome these limitations (Table~\ref{tab:system_comparison}). By combining a robust and modular architecture with simulation-optimized mechanisms, QUASAR enables seamless coordination across diverse computational workflows, ranging from quantum mechanical calculations to classical molecular simulations. Unlike conventional systems, which often require extensive manual setup or specialized expertise, QUASAR leverages the knowledge representation and reasoning capabilities of LLMs to automate complex decision-making processes, such as selecting optimal simulation parameters and dynamically adjusting computational strategies. This integration not only enhances usability, allowing researchers to interact with atomistic simulations more intuitively, but also improves flexibility, supporting a wide range of simulation tools and research objectives. By orchestrating LLM-driven reasoning with high-performance computational routines, QUASAR empowers autonomous scientific research, enabling rapid hypothesis testing, efficient exploration of chemical space, and reliable generation of predictive insights that are traditionally time- and expertise-intensive.

\section{System Architecture and Methods}\label{Method}
\begin{figure}[H]
\centering
\includegraphics[width=\linewidth]{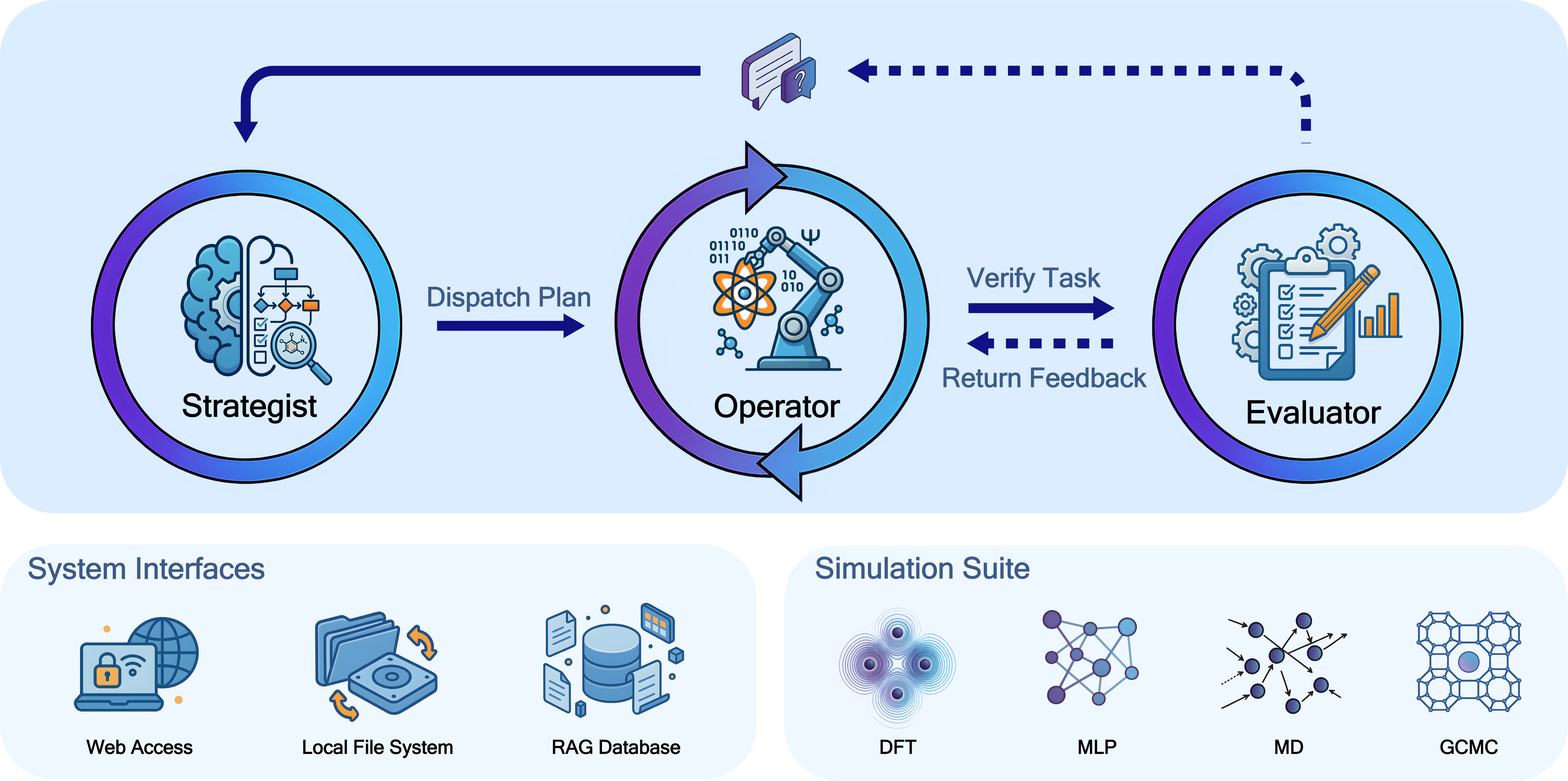}
\caption{Overview of the QUASAR architecture, where dashed lines represent optional feedback paths for iterative refinement.}
\label{fig:QUASAR}
\end{figure}

As shown in Figure~\ref{fig:QUASAR}, QUASAR employs a three-agent architecture built upon LangChain \cite{chase_langchain_2022}. The Strategist interprets user-defined research objectives and decomposes them into scientifically grounded subtasks; the Operator executes these subtasks by interfacing with atomistic simulation software, the local file system, and external resources, handling input preparation, job execution, and analysis; and the Evaluator assesses task completion, returning unsatisfactory results to the Operator for autonomous refinement.

The software stack currently included in QUASAR bridges multiple levels of theory using only open-source software.
This includes Quantum ESPRESSO \cite{giannozzi_advanced_2017} for ab initio DFT calculations, MACE \cite{batatia_mace_2023} for machine-learned potentials (MLP), LAMMPS \cite{thompson_lammps_2022} for classical molecular dynamics (MD), and RASPA3 \cite{ran_raspa3_2024} for grand canonical Monte Carlo (MC) simulations, unified through ASE \cite{hjorth_larsen_atomic_2017} and pymatgen \cite{ong_python_2013} for structure manipulation and analysis.
The multiscale integration allows QUASAR to autonomously chain simulations across length and time scales. As one example, the system can parameterize force fields from quantum calculations and then deploy the resulting force field for large-scale dynamics or adsorption studies that would be intractable at the DFT level. 

In addition to the predefined computational pillars, QUASAR is designed to support dynamic tool integration. Many existing agentic systems rely on hard-coded agent hierarchies and tightly scoped domain-specific tools, which can make their architectures rigid and difficult to extend. Adding new tools to these systems often requires major architecture re-engineering, introducing new specialized agents and functions. QUASAR significantly lowers the barrier to extensibility. For immediate requirements, users can trigger tool installation directly via natural-language prompts. For stable, long-term solutions, developers can bundle tools into the Docker environment with minimal prompt updates. Both pathways significantly reduce integration costs and reinforce the flexibility-oriented design of QUASAR.

\subsection{Robust and Adaptive Planning}
Planning is the most critical stage of the system, as it explicitly defines the Operator's scope of action and establishes the core principles guiding execution. For complex computational chemistry objectives, the direct plan from the Strategist may occasionally overlook key considerations or domain-specific cues. This plan can in turn lead the Operator toward suboptimal or inappropriate actions. For example, in a multi-stage free energy or adsorption workflow, a plan may omit prerequisite equilibration steps which can potentially lead to physically inconsistent or unreliable results. Therefore, to maintain high-quality plan generation, we adopted a double-pass planning mechanism: an initial plan produced by the Strategist is subjected to a second-pass review that explicitly checks missing elements before execution. 

QUASAR also incorporates an iterative feedback loop with optional human-in-the-loop (HITL) integration to address scenarios where ``one-shot'' generation falls short of the scientific objective. After a completed run, researchers may either trigger an automated improvement cycle or provide specific insights. The number of automated iterations can be preconfigured, which allows fully autonomous optimization without requiring continuous manual oversight. This shifts the Strategist from a generative role to a diagnostic one, critically analyzing the previous outputs to propose targeted improvements or corrections.

To further align planning with diverse research objectives, the Strategist is regulated by two user-adjustable parameters: Granularity and Accuracy. Granularity controls task decomposition depth (number of tasks), determining how frequently evaluation and context compression occur. Finer granularity enables more frequent checkpoints but may introduce unnecessary fragmentation for straightforward tasks, while coarser granularity reduces overhead for simpler workflows. Accuracy balances computational cost against precision: ``eco'' mode employs computationally efficient methods for rapid screening, ``standard'' mode provides a balanced approach between speed and accuracy, and ``pro'' mode applies rigorous methods for high-accuracy results at the expense of increased computation time. These parameters enable QUASAR to dynamically adapt its planning and memory architecture to specific research needs. Straightforward tasks such as variable-cell (VC) relaxation in Quantum ESPRESSO benefit from low granularity settings, avoiding redundant task fragmentation and evaluations that would otherwise degrade execution efficiency. This flexibility allows researchers to optimize the balance between thoroughness and computational economy based on their specific objectives.

\subsection{Context and Memory Efficiency}
Tuning context is one of the major challenges in agentic design as it governs the quality and stability of agent orchestration \cite{abou_ali_agentic_2025}. Much like human collaboration, the clarity and volume of the information passed between agents directly dictate the effectiveness of the collective outcome. While recent LLMs feature significantly expanded context windows, they are not infinite. More importantly, analogous to human cognition, as the context length expands, the likelihood of losing focus on the previous conversation increases \cite{du_context_2025}. For instance, as the Operator progresses through a list of tasks, the accumulating context can become excessively large, effectively diluting the model's understanding of completed tasks and escalating the API costs. 

To mitigate this, our system optimizes context sharing at two levels: between individual tasks (task-task) and across distinct execution sessions (run-run). Once the Evaluator verifies the successful completion of a task, it condenses the task context by distilling the valuable actions and outcomes while discarding noise and failed attempts. This not only reduces the overall context length but also strengthens contextual coherence. To support the replanning mechanism discussed earlier, the system additionally maintains long-term memory across runs. In contrast to task-level compression, upon completion of a run the system saves a task-level summary produced by the Evaluator, together with a complete summary generated by the Operator at the final step, to the local filesystem in Markdown format. Enabling seamless connection to prior execution context, QUASAR provides an incremental workflow that drives continuous improvement, and also provides an accessible entry point for human review.

\begin{figure*}[t]
    \centering
    \includegraphics[width=1\linewidth]{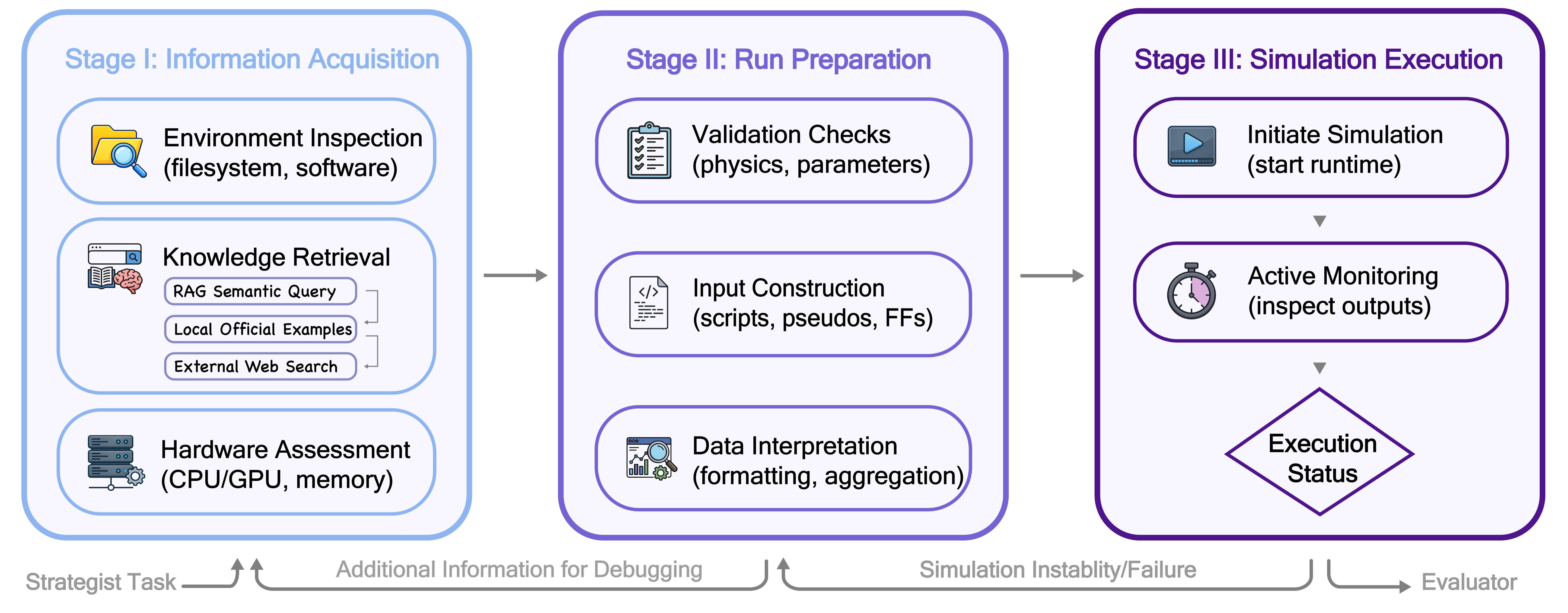}
    \caption{A three-stage pipeline illustrating how the Operator coordinates end-to-end simulation tasks.}
    \label{fig:operator}
\end{figure*}

\subsection{Long Simulation Handling}
The system is designed for reliable operation in real-world scenarios where interruptions from API exhaustion, wall-time limits, or unexpected shutdowns are common. To address this, we implemented persistent state management, ensuring that the full agent state (conversation history and completed steps) is automatically checkpointed after every execution. This allows the system to recover from interruptions without any loss of progress. In addition to the generic checkpointing, QUASAR is specifically optimized for recovering interrupted simulations. When an interrupted simulation is resumed, the system auto-injects an interruption-awareness prompt that guides the Operator to restart from the most recent intermediate state rather than restarting the workflow. For example, during a VC relaxation with Quantum ESPRESSO, the Operator agent will attempt to resume the calculation from the partially relaxed structure instead of initiating a fresh run, significantly reducing redundant computation.

The successful execution and completion of a simulation does not necessarily guarantee scientific correctness or numerical precision. For instance, in software like Quantum ESPRESSO, if a convergence threshold is set too strictly or the system is unstable, the program may reach its default maximum step limit (e.g., 100 steps) and exit without achieving convergence. While the job appears ``Done'' in the output, the simulation has not been completed. In long-duration or computationally expensive simulations, these ``silent failures'' lead to significant waste of time and resources. To address this, we introduce a check-in mechanism to the operator (Stage III in Figure \ref{fig:operator}) that enables active monitoring of the simulation status via a configurable inspection frequency setting. At each check-in, the Operator evaluates intermediate outputs to assess whether the simulation is progressing correctly. If clear issues are detected, such as a lack of DFT convergence, the Operator terminates the run and attempts to reconstruct or optimize the input parameters or other related factors.

\subsection{Hybrid Knowledge Retrieval}
QUASAR employs a domain-specific Retrieval-Augmented Generation (RAG) database, indexing documentation and source code to mitigate hallucinations arising from limited or outdated LLM knowledge. However, simulation engines such as LAMMPS and Quantum ESPRESSO rely on highly specialized and unique input syntaxes. Their official example scripts and input files typically contain very minor natural language comments and instead rely on a separate overview README file for explanation. This sparse semantic context makes embedding-based retrieval challenging, as the technical syntax and domain-specific parameters lack the descriptive text that vector search relies upon. To overcome this limitation, QUASAR exposes a local repository of example input files directly to the Operator. Rather than relying solely on semantic embeddings, the Operator leverages its reasoning capabilities to navigate to the provided directories, interpret filenames, and parse README files to identify relevant reference examples. 

Additionally, QUASAR implements a hierarchical knowledge-access shown in Stage I in Figure \ref{fig:operator} via the Operator's system prompt to further enhance the retrieval efficiency. When the Operator has high confidence in the required simulation physics and input syntax, it proceeds autonomously using its internal knowledge. If uncertainty arises, it is guided to first perform semantic similarity matching via RAG to relevant documentation or code examples. When a RAG query is insufficient or ambiguous, the system uses the logical inference approach, in which the Operator systematically explores the provided example repositories and infers appropriate simulation patterns. Only as a final fallback does the Operator expand its search space to external web resources. 

By escalating from quick-access internal data to resource-heavy external searches only when necessary, QUASAR allows the system to reliably support complex simulation setups and debugging tasks, ensuring higher rates of successful simulation execution while reducing unnecessary action overhead. The practical importance of this hierarchy is evident when external knowledge retrieval is fully disabled. Without it, the Operator consistently fails to construct valid input files for RASPA3, likely due to the relatively sparse representation of RASPA3 in LLM training data. This outcome demonstrates that the external knowledge layer provides meaningful support beyond the inherent model knowledge alone.

\subsection{Containerized Environment and HPC Optimization}
Usability and security are two critical considerations in the design of QUASAR. To ensure seamless deployment across diverse environments, QUASAR provides cross-platform Docker containers pre-packaged with the suite of computational tools stated previously. This includes specialized images optimized for CUDA and ROCm to facilitate hardware-accelerated inference (e.g., for MACE models). Moreover, from a security perspective, the containerized nature inherently prevents catastrophic actions such as deletion of system files. QUASAR can also operate completely offline and supports integration with locally hosted open-source LLMs, ensuring that sensitive or confidential data remains secure.

QUASAR is deeply optimized for high-performance computing (HPC), and its Docker containers are designed to run natively under Singularity/Apptainer on HPC systems. Its non-interactive execution mode paired with a built-in restart mechanism, allows entire workflows to be submitted and reliably executed as standard batch jobs. The system currently does not support agent-driven HPC job submission, as it is designed to prioritize broad compatibility across diverse computational environments. The variations in HPC schedulers, configurations, and site-specific policies make it difficult to reliably abstract them into a single generalized interface. Therefore, for specialized deployments targeting a single fixed HPC environment, additional facility-specific submission tooling could be incorporated, but this falls outside the current design scope of QUASAR.

\subsection{User Interfaces}
To connect the system backend with end users, QUASAR provides a Command Line Interface (CLI) with essential control and monitoring capabilities. The CLI permits real-time tracking of agent progress and provides granular visibility into execution details, including the generated plan and code snippets. It also supports a robust restart mechanism, allowing users to resume execution after interruptions. For advanced use cases, an optional web-based graphical user interface (GUI) is available through licensing agreements, offering streamlined monitoring, fine-grained execution control, and visualization.

\section{Results}\label{Results}
We evaluated QUASAR using a three-tiered benchmark suite (Table~\ref{tab:benchmark_overview}) with the \texttt{gemini-3-flash\-preview} LLM. All tiers were executed multiple times to demonstrate the consistency of QUASAR’s results. Detailed metadata, including hardware details and token usage, is provided in the Supporting Information.

\begin{table}[h!]
    \centering
    \caption{Overview of the three-tiered benchmarking suite used to validate QUASAR. Row headers are highlighted in grey to distinguish tiers. For Tier I and II cases, the results are presented as either an average or a uniform outcome based on three runs. Tier III cases were executed twice to verify consistency.}
    \label{tab:benchmark_overview}
    
    \vspace{0.3cm}
    \small
    \renewcommand{\arraystretch}{1.4} 
    
    \begin{tabularx}{\textwidth}{
        >{\bfseries\raggedright\arraybackslash}p{2.8cm} 
        X 
        >{\raggedright\arraybackslash}p{3.8cm}
    }
        \toprule 
        Test Case & \textbf{Prompt} & \textbf{Result} \\
        \midrule

        \multicolumn{3}{l}{\textcolor{black!60}{\textit{\textbf{Tier I: Task Execution}}}} \\
        \addlinespace[0.2em]
        
        K-point Convergence 
        & Calculate the k-point density to converge bulk Cu energy to 1 meV/atom.
        & $N_k = 14 \times 14 \times 14$ \newline \textcolor{gray}{True} \\
        
        NPT Equilibration 
        & Calculate the density of water at 298 K and 1 bar.
        & $\bar{\rho} = 0.99787$ g/cm$^3$ \newline \textcolor{gray}{Ref: 0.99709 g/cm$^3$ \cite{informatics_nist_nodate}} \\
        
        Helium Void Fraction 
        & Calculate the helium-accessible void fraction  for IRMOF-1.cif at 298~K.
        & $\bar{\phi}_{He} = 0.8177$ \newline \textcolor{gray}{Ref: approx. 0.80 \cite{noauthor_raspa_nodate}} \\
        \midrule

        \multicolumn{3}{l}{\textcolor{black!60}{\textit{\textbf{Tier II: Workflow Orchestration}}}} \\
        \addlinespace[0.2em]

        Band Gap 
        & Calculate the electronic band gap for bulk nickel oxide.
        & $\bar{E}_g = 4.26$ eV
        \textcolor{gray}{Ref: 4.0 eV \cite{hufner_electronic_1994}} \\
        
        Adsorption Isotherm 
        & Calculate the adsorption isotherm for CO$_2$ in UiO-66 at 298~K.
        & $\bar{n}(10\,\mathrm{bar}) = 6.22$ mmol/g \newline \textcolor{gray}{Ref: approx. 5 - 7 mmol/g \cite{jajko_carbon_2021}} \\
        
        Melting Point 
        & Calculate the melting point of aluminum.
        & $\bar{T}_m = 928.70$ K \newline \textcolor{gray}{Ref: 933.45 K \cite{informatics_nist_nodate}} \\
        \midrule

        \multicolumn{3}{l}{\textcolor{black!60}{\textit{\textbf{Tier III: Frontier Research}}}} \\
        \addlinespace[0.2em]

        Photocatalyst Screening 
        & Determine which 5\% La-doped ATaO$_3$ perovskite (A = Li, Na, K) exhibits the best photocatalytic degradation performance against methyl orange under UV irradiation.
        & NaTaO$_3$: 5\% La \\
        
        Gas Separation 
        & Determine which COF maximizes the Xe/Kr adsorption selectivity at 298 K. The experimental CIF structures are provided.
        & COF-X-Br \\
        
        Virtual MOF Assessment 
        & Determine the CO$_2$ adsorption performance and mechanical properties of the provided MOF.
        & Non-porous ($q \approx 0$) \\
        \bottomrule 
    \end{tabularx}
\end{table}

\subsection{Tier I. Task Execution}
To verify the baseline usability of the system, we selected three simple representative tasks spanning distinct methodological archetypes. These foundational tests are designed to confirm that QUASAR can reliably execute well-defined, single-step calculations that form the building blocks of more complex workflows. While individually straightforward, each task requires correct parameter selection, appropriate software invocation, and accurate interpretation of outputs.

The first task involves DFT k-point convergence, which ensures reliable Brillouin-zone sampling for electronic structure simulations and is a prerequisite for any production-quality periodic calculation. The second task addresses $NPT$ ensemble equilibration in molecular dynamics, which is essential for obtaining thermodynamically consistent densities before production simulations and validates the capacity of QUASAR to configure and monitor classical simulation protocols. The third task focuses on helium void-fraction analysis, a Monte Carlo characterization method used to quantify accessible porosity in porous materials for gas storage and separation applications. Together, these three tasks span quantum mechanical, classical mechanics, and Monte Carlo methods, providing a representative cross-section of the computational toolkit commonly employed in materials science research.

\subsection{Tier II. Workflow Orchestration}
Moving beyond individual calculations, Tier II evaluates the ability of QUASAR to decompose high-level scientific questions into coherent, multi-step computational workflows.

The first task involves calculating the electronic band gap of bulk nickel oxide, which challenges the system to handle strongly correlated transition metal oxides. This requires optimizing the crystal structure, correctly initializing the Type-II antiferromagnetic ordering, and applying an appropriate Hubbard $U$ correction (DFT+$U$) or hybrid functional to accurately reproduce the band gap.

The second task focuses on generating \ce{CO2} adsorption isotherms for UiO-66 at 298~K. To produce a full isotherm, the system must orchestrate a series of Grand Canonical Monte Carlo (GCMC) simulations, which involves selecting appropriate force fields, preparing the simulation cell, and systematically sweeping through a range of chemical potentials (pressures) to map the isotherm.

The final task in this tier requires determining the melting point of aluminum, testing the ability of QUASAR to characterize phase transitions using molecular dynamics methods. The agent must autonomously infer and correctly sequence all prerequisite computational steps, such as system equilibration and phase initialization prior to executing the melting-point calculation. It then needs to select a reliable approach such as the two-phase coexistence method or temperature curves to accurately reproduce the solid–liquid transition temperature.

\subsection{Tier III. Frontier Research}
While the Tier II tasks demonstrate strong potential for real-world applications, they are well-documented in the literature and possibly represented in LLM training data. Tier III therefore targets recent or open research problems that lack established results, thereby testing the ability of the system to reason and explore without reliance on known solutions.

We tasked QUASAR with screening 5\% La-doped \ce{ATaO3} perovskites (A = Li, Na, K) for the photocatalytic degradation of methyl orange under UV irradiation. This complex query requires the agent to compute and align band edges relative to relevant redox potentials, assess the impact of doping on the optical gap, and evaluate defect formation energies. This task is based on recent published work from our lab \cite{matthews_doping_2025}.

The system was further challenged to identify which candidate covalent organic framework (COF) maximizes selectivity for xenon over krypton (Xe/Kr) at 298~K, from a selection of two unpublished COF structures. Research currently under review features these structures and they were experimentally verified. This task demands a comparative screening workflow that captures subtle host–guest interactions distinguishing these two noble gases, necessitating precise force-field selection and competitive adsorption simulations.

Finally, we provided the system with a novel material structure generated via latent diffusion that is not yet experimentally synthesized \cite{simkus_mofasa_2025}. We tasked the agent with simulating its mechanical properties and \ce{CO2} adsorption capacity. This stress test evaluates robustness in handling potentially unstable or unoptimized geometries, requiring structural integrity validation (e.g., via elastic tensor calculations) before proceeding to functional property prediction.

\section{Discussion}
The performance of QUASAR in Tier I and Tier II highlighted both the strengths and limitations of reasoning-driven computation. The system achieved accurate results with efficiency and rigor across all cases in these tiers It is worth noting that in the Tier II NiO band gap calculation case, the system failed to obtain an accurate value for the first run and required one or two further auto-improvement run contexts to recognize the need to switch to the higher-fidelity Heyd–Scuseria–Ernzerhof (HSE) method. This behavior likely reflects the strong training priors of the LLM, as DFT+$U$ is overwhelmingly reported as the default approach for NiO in the literature. However, with the auto-improvement mechanism, QUASAR was able to escape this ``local optimum'' and progress toward a more accurate solution, illustrating the value of iterative, incremental approaches to goal-directed computation.

Building upon this foundation, the Tier II analysis underscores the importance of Tier III testing, which evaluates not only the robustness of trained routines but also the capacity for LLMs to apply learned knowledge in novel scenarios. In this tier, the final photocatalyst screening results correctly recreated the published result \cite{matthews_doping_2025}, and the gas separation and virtual MOF cases are validated as physically reasonable and methodologically reliable (see Supporting Information). Importantly, the performance of QUASAR in Tier III challenges the conventional assumption that materials research requires HITL guidance. These tasks did not undergo auto-improvement and the system performed end-to-end screening, method selection, and analysis, suggesting that agentic AI systems can independently execute complex scientific workflows with minimal or no human intervention. Our results highlight that QUASAR, and similar systems, may exhibit a level of autonomy that approaches that of expert researchers, while the boundaries of this capability remain to be fully determined. It is also worth noting that QUASAR achieved reasonable and high-quality results across all three benchmark tiers using a lightweight , eco-friendly model \texttt{gemini-3-flash-preview}. This suggests that frontier models, such as \texttt{gemini-3-pro-preview} with enhanced reasoning capabilities, may not only pass the proposed benchmarks but do so with greater efficiency and rigor. 

The non-deterministic nature of LLMs should also be addressed as it may compromise the reproducibility of scientific results. QUASAR is designed to mimic the behavior of a human computational researcher. Just as no two researchers approach a computational task in exactly the same way, QUASAR may employ different strategies or parameters across independent runs when presented with the same query. Crucially, however, the three repeated trials across both Tier I and Tier II benchmark cases reflect that the variability does not undermine the overall goal of the user request. Moreover, QUASAR preserves all inputs and step-by-step explanations throughout each run, which means it is straightforward to reproduce a prior trajectory by supplying this recorded evidence, analogous to how human researchers maintain detailed logs and methods to ensure their work is reproducible. We also observe that when QUASAR fails to complete a task correctly, the Evaluator detects the failure and issues a correction. In some instances, the system bypasses the problematic step and continues execution. This behavior mirrors the reality of human research, where initial attempts are not always successful. The failure recovery of QUASAR and iterative improvement mechanism is specifically designed to address such scenarios, enabling the system to identify, learn from, and overcome failures in subsequent runs.

Ultimately, these findings indicate that QUASAR is not merely an automation layer over broader simulation tools, but a fundamentally different paradigm for conducting atomistic research. Rather than encoding fixed workflows or predefined procedural pipelines, QUASAR demonstrates effective reasoning-driven orchestration, where planning, execution, evaluation, and refinement emerge from coordinated agent cognition. This shift is critical: the future direction of agentic simulation frameworks may no longer rely on heavily engineered, rigid infrastructures, but instead focus on expanding the boundaries of autonomous reasoning, where systems adaptively construct, evaluate, and refine their own computational strategies in response to evolving scientific objectives. The three-tiered benchmark demonstrates strong task flexibility across diverse simulation domains, though its generalization beyond the evaluated suite requires further study. It should also be noted that, because the system is fundamentally dependent on LLM capabilities, performance remains bounded by the underlying domain knowledge and in-context reasoning ability of the chosen model. When a model with incomplete or imprecise knowledge is given the authority to direct simulations, errors may propagate undetected, wasting computational resources or producing misleading results that appear superficially plausible. Unlike conventional software failures that manifest as explicit exceptions, LLM-induced errors often present as structurally coherent but numerically or physically unsound outputs, surfacing only under expert scrutiny.

The rise of AI-driven tools like QUASAR also raises important philosophical and practical questions about how researchers should responsibly engage with computational chemistry. While QUASAR can significantly reduce cognitive load for running simulations and theoretical validation, this efficiency comes with the risk of skill erosion that users who rely solely on the software may lose a deep understanding of the underlying methods and concepts. Yet this shift may also reflect an evolution in the field, where the delegation of routine computational tasks to intelligent tools frees researchers to rethink methodologies, challenge theoretical assumptions, and drive conceptual innovation that software alone cannot achieve. At a stage where AI autonomy remains the subject of ongoing debate, the most effective path forward is to benchmark these tools widely and comprehensively to establish their reliability across diverse contexts. For now, human oversight remains necessary to detect errors introduced by the agent and to ensure that outputs are physically plausible.

\section{Summary and Outlook}\label{Conclusion}
In this work, we introduced QUASAR, an open-source, portable, and universal atomistic computation system designed to transition agentic AI from simple demonstrations and basic task handling toward production-grade scientific discovery. The flexible architecture of QUASAR supports further customization with minimal effort, such as the seamless integration of additional tools.

Through our Tier III benchmarking, QUASAR demonstrates that LLMs, when properly orchestrated, are capable of executing sophisticated computational chemistry tasks with a level of scientific rigor comparable to that of human researchers. This work serves as a proof of concept that modern AI has the potential to automate the majority of standard computational workloads, freeing the scientific process from the bottlenecks of manual execution. This positions QUASAR as a prototype for human–AI collaborative workflows in computational chemistry, where AI handles routine orchestration while humans provide scientific direction, judgment on ambiguous cases, and validation of results. 

Beyond the \texttt{gemini-3-flash-preview} evaluated in this study, a diverse array of frontier LLMs, along with specialized small-scale models tailored for computational chemistry tasks, continue to emerge \cite{shi_mdagent2_2026}. Given this diversity, selecting the best-performance model remains challenging when relying solely on success rates. It is imperative that future directions focus on developing comprehensive benchmarks (beyond the limited set employed in this study) to evaluate not only whether scientific objectives are achieved, but also the broader capabilities underlying the success. These benchmarks are necessary to fairly test the strategies employed, and the efficiency of decision-making across different models and agentic systems.

Our study also offers a perspective on the necessary evolution of the current scientific process as AI capabilities are reaching unprecedented heights. As execution, debugging, and workflow construction become automated, the primary role of human researchers shifts toward conceptual framing and theoretical innovation. This transition does not displace scientists but reallocates cognitive responsibility, enabling human researchers to prioritize scientific scope and hypotheses. We envision that AI may fundamentally reshape the structure and dynamics of the research lifecycle in the near future.

\section*{Supporting Information}
All code and data for this research are available on Zenodo: \href{https://doi.org/10.5281/zenodo.18409876}{10.5281/zenodo.18409876}. Additionally, the source code is available on GitHub \href{https://github.com/fengxuyy/QUASAR}{\faGithub}, and the docker images for QUASAR deployment are hosted on DockerHub \href{https://hub.docker.com/r/fengxuyang/quasar}{\faDocker}.

\section*{Acknowledgements}
J.D.E. is the recipient of an Australian Research Council Discovery Early Career Award (project number DE220100163) funded by the Australian Government. The Phoenix HPC service at the Adelaide University is acknowledged for providing high-performance computing resources. This research was supported by the Australian Government's National Collaborative Research Infrastructure Strategy (NCRIS), with access to computational resources provided by Pawsey Supercomputing Research Centre through the National Computational Merit Allocation Scheme. We thank Mohamad Moosavi (University of Toronto) and his team for thought-provoking discussion.

\end{document}